\documentclass{PoS}

\usepackage{graphicx}
\usepackage{amsmath}
\usepackage{caption}
\usepackage{subcaption}
\usepackage{algorithmicx,algpseudocode}
\usepackage{algorithm}
\usepackage{placeins}

\def\Tr{\operatorname{Tr}}
\def\Var{\operatorname{Var}}
\DeclareMathOperator*{\sumsum}{\sum\sum}

\title{Algorithms for Disconnected Diagrams in Lattice QCD}

\ShortTitle{Algorithms for Disconnected Diagrams in Lattice QCD}

\author{\speaker{Arjun Singh Gambhir}\\
        Department of Physics, The College of William and Mary, Williamsburg, VA, 23185, U.S.A.\\
        Thomas Jefferson National Accelerator Facility, Newport News, VA 23606, U. S. A.\\
        E-mail: \email{asgambhir@email.wm.edu}}

\author{Andreas Stathopoulos\\
        Department of Computer Science, The College of William and Mary, Williamsburg, VA, 23185, U.S.A.\\
        E-mail: \email{andreas@cs.wm.edu}}  

\author{Kostas Orginos\\
        Department of Physics, The College of William and Mary, Williamsburg, VA, 23185, U.S.A.\\
        Thomas Jefferson National Accelerator Facility, Newport News, VA 23606, U. S. A.\\
        E-mail: \email{kostas@jlab.org}}
        
\author{Boram Yoon\\
        Los Alamos National Laboratory, Los Alamos, NM, 87545, U.S.A.\\
        E-mail: \email{boram@lanl.gov}}
        
\author{Rajan Gupta\\
        Los Alamos National Laboratory, Los Alamos, NM, 87545, U.S.A.\\
        E-mail: \email{rg@lanl.gov}} 
        
\author{Sergey Syritsyn\\
        Department of Physics and Astronomy, Stony Brook University, Stony Brook, NY 11794, U.S.A.\\
        E-mail: \email{sergey.syritsyn@stonybrook.edu}}                                

\abstract{Computing disconnected diagrams in Lattice QCD (operator
  insertion in a quark loop) entails the computationally demanding
  problem of taking the trace of the all to all quark propagator. We
  first outline the basic algorithm used to compute a quark loop as
  well as improvements to this method. Then, we motivate and introduce
  an algorithm based on the synergy between hierarchical probing and
  singular value deflation. We present results for the chiral
  condensate using a 2+1-flavor clover ensemble and compare estimates
  of the nucleon charges with the basic algorithm.}

\FullConference{34th annual International Symposium on Lattice Field Theory\\
		24-30 July 2016\\
		University of Southampton, UK}

\begin{document}

\section{Introduction}

Lattice QCD calculations of hadron structure observables such as form
factors require computing ``disconnected diagrams". These are diagrams
in which a quark loop with operator insertion is coupled to the nucleon correlator
via gluons only. Such diagrams are required for isoscalar and
strange form factors of the nucleon.

The computation of the quark loop requires the trace of the inverse of
the Dirac operator over spin, color, and spatial dimensions. An exact
computation of this trace would be proportional to the square of the
lattice volume, instead, stochastic methods are employed. The basic
technique is a Monte Carlo (MC) averaging over matrix quadratures: the
Hutchinson trace \cite{Hutchinson_90, Dong:1993pk}.
\begin{equation}
t(A^{-1}) \approx \frac{1}{s}\sum_{j=1}^{s}z_j^\dagger A^{-1}z_j
\label{trace}
\end{equation}
where the $z_j$ are random noise vectors with elements that satisfy
$E(z_j(k)z_j(k')) = \delta_{kk'}$. In particular, for the case of
$\mathbb{Z}_4$ noise ($\pm 1,\pm i$ uniformly distributed elements),
the variance of this estimator is
\begin{equation}
\Var({\rm Tr}(A^{-1}))=\left( \|A^{-1}\|_F^2-\sum_{i=1}^{N}|A^{-1}_{i,i}|^2\right)=\sumsum_{i\ne j}^{N}|A^{-1}_{i,j}|^2.
\label{variance}
\end{equation}

From equation \ref{variance}, it is obvious that large off diagonal
elements slow down the converge of the MC estimator. Many improvements
to this basic estimator have been proposed and implemented in the
Lattice community. Recent advances include dilution and eigenvalue
deflation \cite{Neff:2001zr, Foley:2005ac, Babich:2007jg}.

\section{Hierarchical Probing and Deflation}

It is well known that elements of $A^{-1}$ between two points decay as a
function of geometric distance on the lattice. Therefore, an improved
estimator with smaller variance than that shown in eq.~\ref{variance}
should take advantage of this feature and remove or reduce the
contributions from the largest off diagonal matrix elements. One
possible idea it to use probing instead of MC with large numbers of $z_j$
\cite{Tang_Saad_traceInv}. This is accomplished by graph coloring the
sparsity pattern produced by the Dirac matrix to generate special
vectors that expose the trace. Probing removes local matrix elements
up to a specified distance, thereby eliminating the largest components
of error. We refer to this algorithm as classical probing (CP).

A drawback of CP is that probing vectors for a particular coloring
distance are not reusable at higher distances. A solution to
this is to first generate $m$ probing vectors based on known
structure, such as spin, color, or time. Then, $s$ random vectors are
created that follow the pattern produced by each probing vector,
resulting in $m\times s$ total vectors. This method is called dilution
\cite{Babich:2011, Morningstar_Peardon_etal_2011}.

Hierarchical Probing (HP) expands upon CP and dilution by probing in a
nested way so that the vectors at a particular probing distance are
reused in all higher coloring distances. The HP basis is generated by
permutations of the Hadamard matrix for lattices with purely even
powers. HP annihilates the largest off diagonal elements of $A^{-1}$
in an incremental way, and has been shown to give an order of
magnitude improvement over random noise for strange quark
contributions \cite{Stathopoulos:2013aci}.

Permutations of the Hadamard matrix produce a deterministic algorithm,
however the estimator can be made unbiased by performing an
element-wise product with a random $\mathbb{Z}_4$ vector. Therefore, 
HP may be viewed as an algorithm that generates a nested probing
basis from any starting vector.

For light quark masses, $A^{-1}$ becomes dominated by low lying eigenvectors 
and the off-diagonal elements decay slowly. We, therefore, supplement HP with deflation, 
removal of low modes from $A$ via a projection, to recover 
the strong decay structure, i.e., we calculate the trace of the deflated matrix 
with HP and correct it by adding to it the exact low mode
contribution. In our implementation, we deflate $A$ using the 
singular value decomposition (SVD) instead of the typical
eigendecomposition. The effects of SVD deflation are studied
theoretically and numerically in \cite{Gambhir:2016uwp}. The improved
deflated HP estimator is given by
\begin{equation}
\operatorname{Tr}(A^{-1}) \approx \frac{1}{s}\sum_{j=1}^{s}\left(z_j^\dagger A^{-1}z_j-z_j^\dagger V\Sigma^{-1}U^\dagger z_j\right)+\operatorname{Tr}(V\Sigma^{-1}U^\dagger ),
\label{deflatedtrace}
\end{equation}
where $U$, $\Sigma$, and $V$ are the low lying singular triplets of
$A$. 

In order for the total algorithm to be cost-effective, deflation must
be extremely efficient. Possible algorithms for determining the low
lying subspace are Lanzcos or eigCG \cite{Stathopoulos:2007zi}. We
chose instead the PReconditioned Iterative MultiMethod Eigensolver
(PRIMME) \cite{Stathopoulos06primme:preconditioned} library,
accelerated with a multigrid preconditioner \cite{Babich:2010qb}, to
solve for the eigenpairs $\Lambda, V$ of $A^\dagger A$. This allowed
us to efficiently generate 1,000 lowest singular triplets of $A$. We
stress the importance and computational challenge of this problem due
to the high density of the low lying spectrum of the Dirac
matrix. Details of our procedure can be found in
\cite{Gambhir:2016uwp}.

With the low mode space calculated, 
the ``Deflated Trace Estimation Algorithm'' is as follows: 

\begin{algorithm}
\caption{$Trace = $ deflatedHP$(A)$ }
\begin{algorithmic}[1]
\State $[\Lambda, V] = \operatorname{PRIMME}(A^\dagger A)$ 
\State $T_D = \Tr(\Lambda^{-1}V^\dagger A^\dagger V)$; $T_R = 0$
\State $z_0=\operatorname{randi}([0,1], N, 1)$; $z_0=2z_0-1$
\For{$j=1:s$}
\State $z_h = $ next vector from Hierarchical Probing or other scheme
\State $z = z_0 \odot z_h$
\State Solve $Ay = z$
\State $T_R = T_R + z^\dagger y - z^\dagger V \Lambda^{-1} V^\dagger (A^\dagger z)$
\EndFor
\State $Trace = T_R/s + T_D$
\end{algorithmic}
\label{Alg:main}
\end{algorithm}

\FloatBarrier

\section{Numerical Experiments}

\subsection{Variance and Speedup Tests}

We computed $\Tr A^{-1}$ to estimate the variance reduction and
speedup of the combined, deflated HP, algorithm. These tests were
performed on $32^3\times 64$ $2+1$-flavor Clover-on-Clover fermion
lattices with spacing $a=0.0081 \ fm$ and pion mass $m_\pi=312 \ MeV$.

\begin{figure}[h]
  \centering
  \subcaptionbox{Variance of $\Tr(A^{-1})$ Variance\label{m239VarianceLog}}{\includegraphics[width=0.49\textwidth]{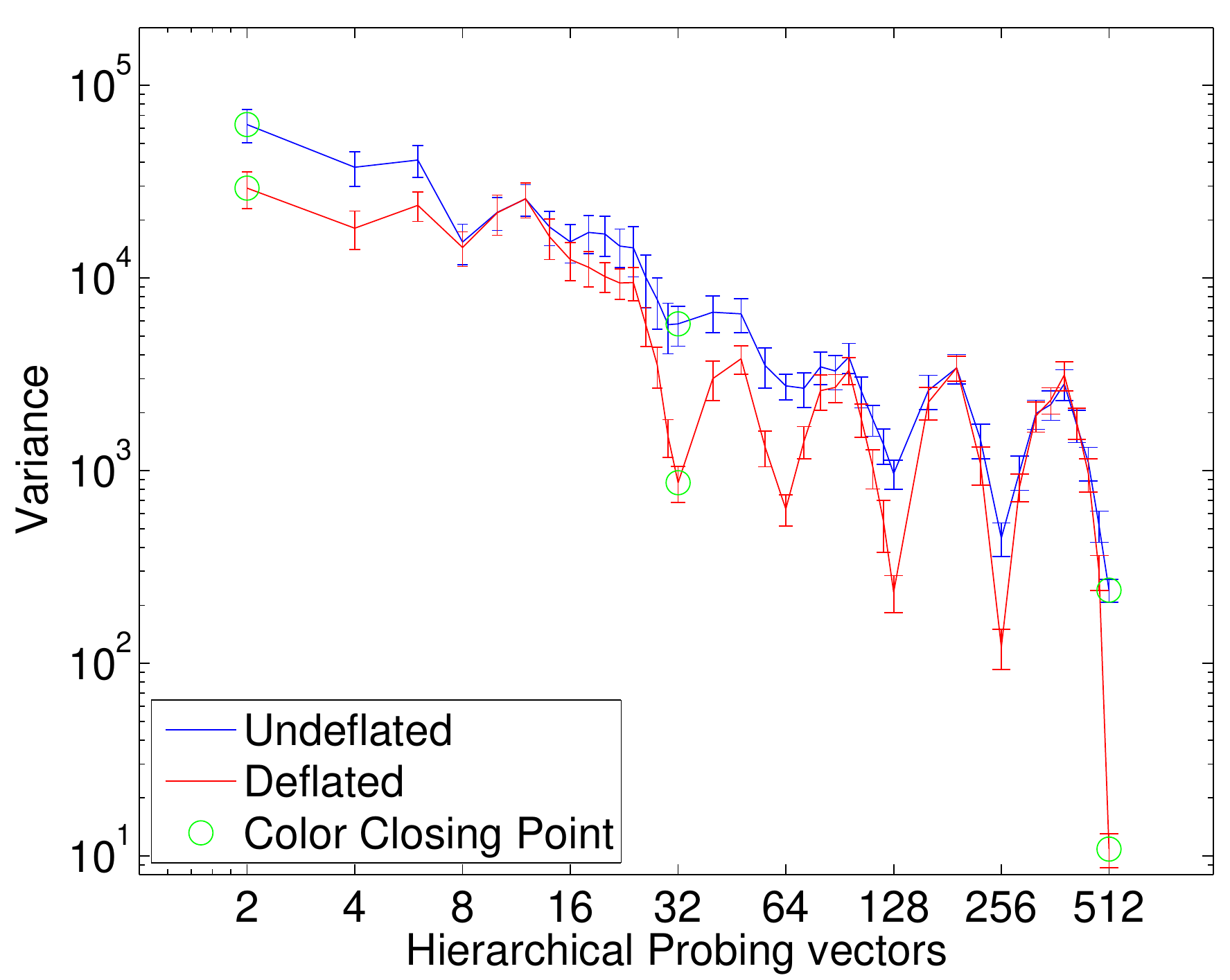}}\hspace{0em}%
  \subcaptionbox{Speedup of $\Tr(A^{-1})$ Variance\label{m239SpeedUp}}{\includegraphics[width=0.49\textwidth]{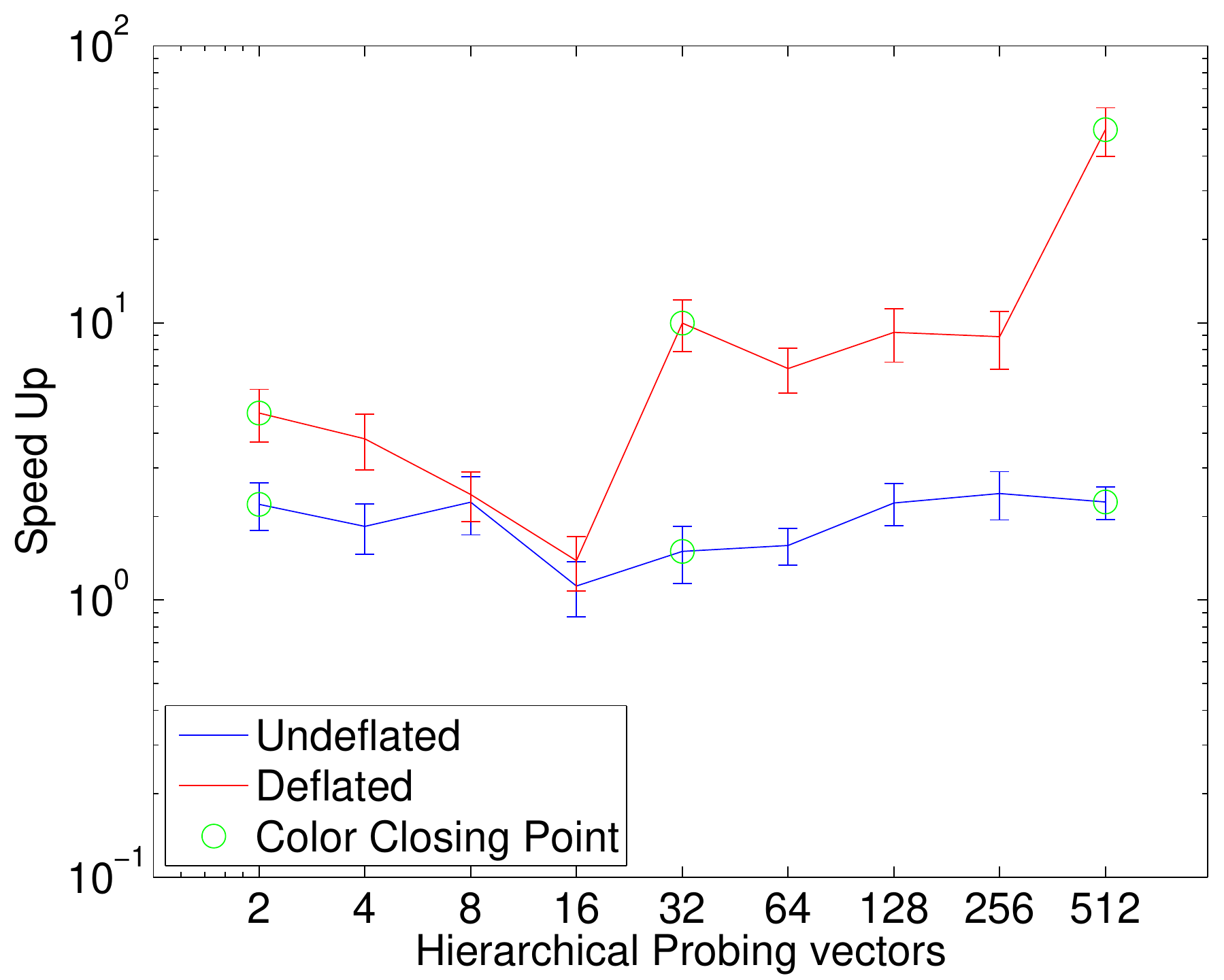}}
  \caption{The left plot shows the variance of the deflated HP trace
    estimator. Speedup in comparison to pure $\mathbb{Z}_4$ noise is
    shown on the right. The points 2, 32, and 512 are the color
    closing points of the nested basis and give the highest variance
    reduction, and conversely the best speedup. The errors on both
    figures were computed with jackknife resampling over 40
    $\mathbb{Z}_4$ noise vectors.}
\end{figure}

Figure~\ref{m239VarianceLog}, shows that deflation reduces variance of
HP by a factor of 20 at the third color closing point.  The speedup in
Fig.~\ref{m239SpeedUp} is given by
$R_s=\frac{V_{stoc}}{V_{hp}(s)\times s}$ where $V_{stoc}$ is the
variance of the purely stochastic MC estimator, and $V_{hp}(s)$ is the
variance of deflated HP. The factor of $s$ in the denominator
normalizes against pure noise since error from the MC estimator falls
off as $\sqrt{\frac{V_{stoc}}{s}}$. Fig.~\ref{m239SpeedUp} shows that
at light quark masses HP alone yields only a factor of 2-3 speedup
over the basic method (blue spline). Augmented with deflation, HP gives a
factor of 60 speedup at the third color closing point.

\subsection{Synergy between Deflation and HP}

\begin{figure}[h]
\centering
\includegraphics[width=0.6\textwidth]{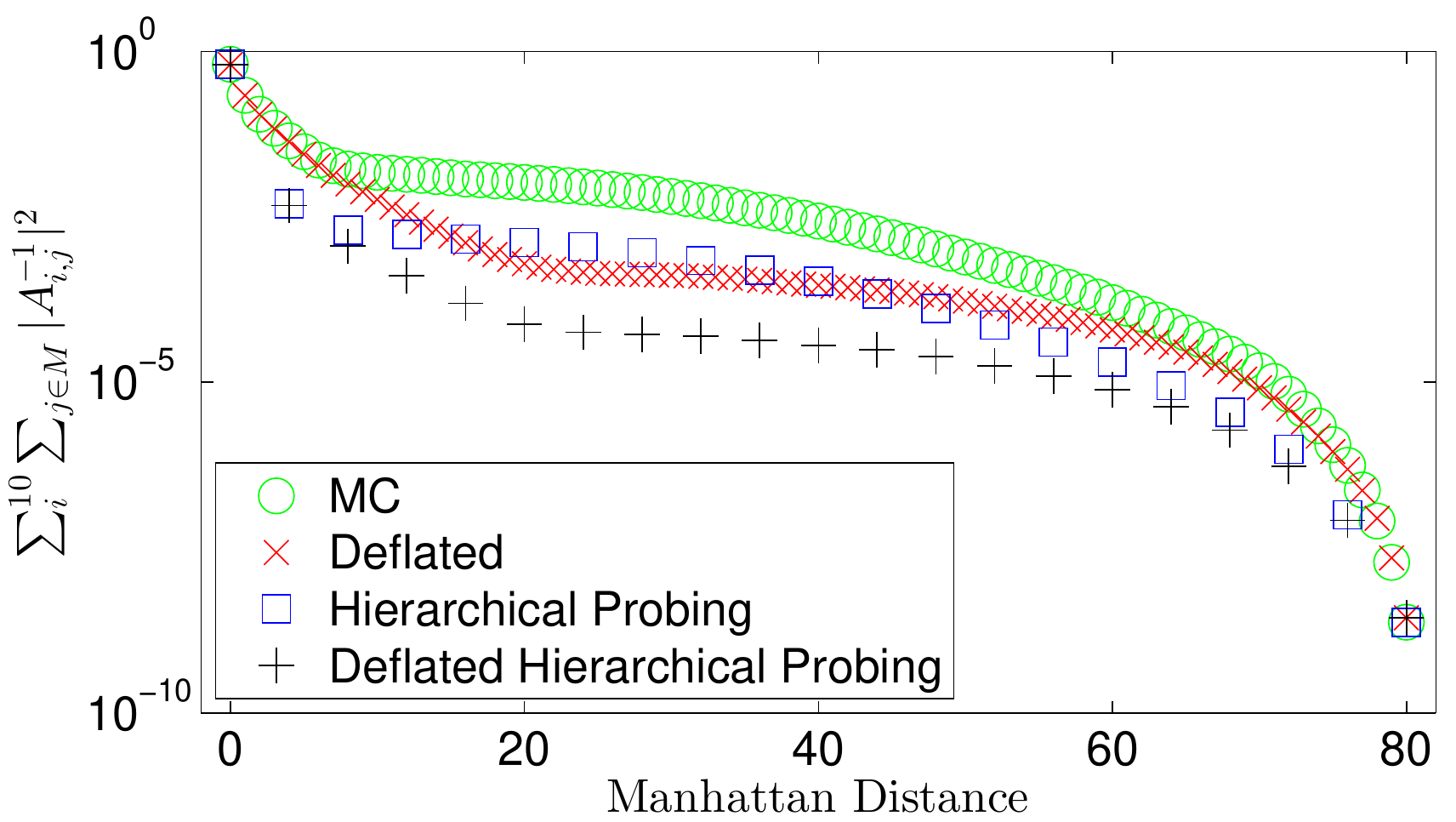}
\caption{Deflation alone includes the full space of 1,000 singular
  triplets. HP alone is the application of 32 hadamard
  vectors. Deflated HP is the combination of both methods.}
\label{m250Taxi}
\end{figure}

To understand why these two methods, deflation and HP, work well
together, we sample 10 randomly selected rows $i$ of the Dirac matrix used above, $A^{-1}_{i,j}$, as
a function of their Manhattan distance $M$ along the lattice ($j$). We
exhibit results for four cases in Fig.~\ref{m250Taxi}: an MC
estimator, deflation alone, HP alone, and deflation combined with HP.
Data in Fig.~\ref{m250Taxi} show that HP reduces the off-diagonal elements of $A^{-1}$ 
at a small Manhattan distance faster (blue squares lie below green
circles and red crosses at $M\sim 4$). From $M=10$ to $M=50$,
deflation is more effective (red crosses lie below green circles and
blue squares). Together, these techniques yield a highly efficient
trace estimation algorithm (black plus symbol).

\subsection{Nucleon Charges} 

\begin{figure}[!ht]
\centering
\subcaptionbox{$g_A$ 100 cfgs\label{gA_100_comparison}}
{\includegraphics[width=0.45\textwidth]{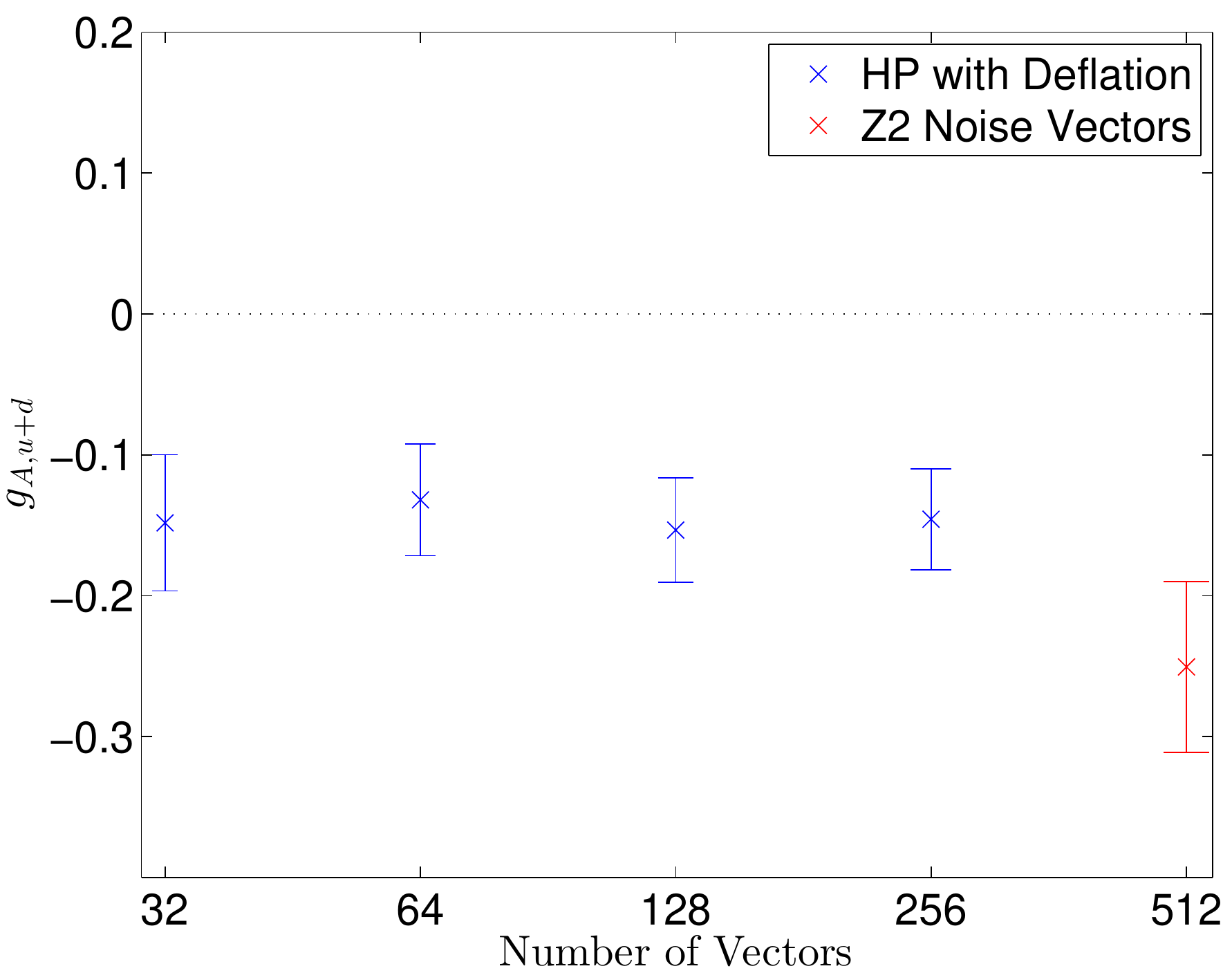}}
\subcaptionbox{$g_S$ 100 cfgs\label{gS_100_comparison}}
{\includegraphics[width=0.45\textwidth]{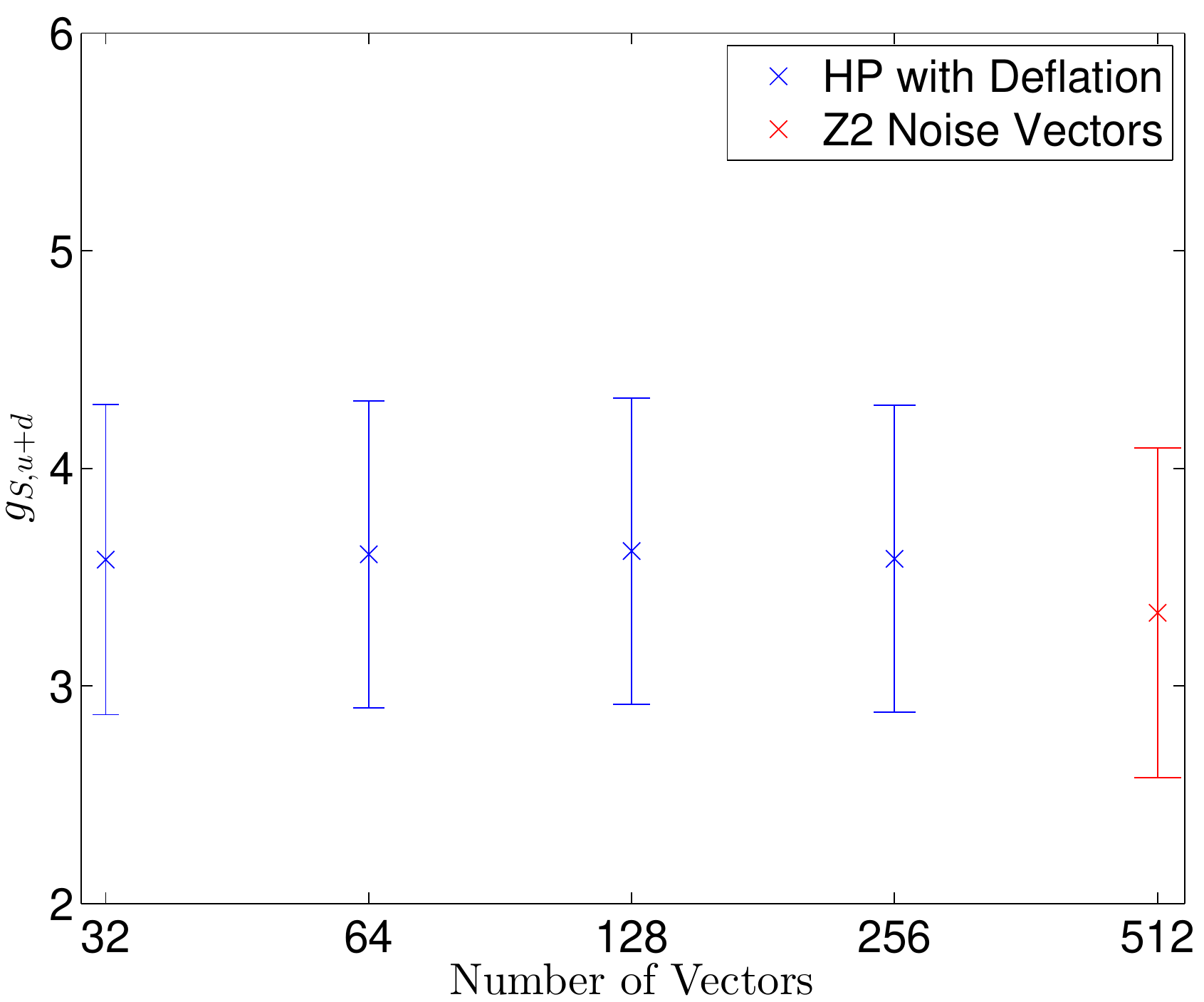}}
\subcaptionbox{$g_T$ 100 cfgs\label{gT_100_comparison}}
{\includegraphics[width=0.45\textwidth]{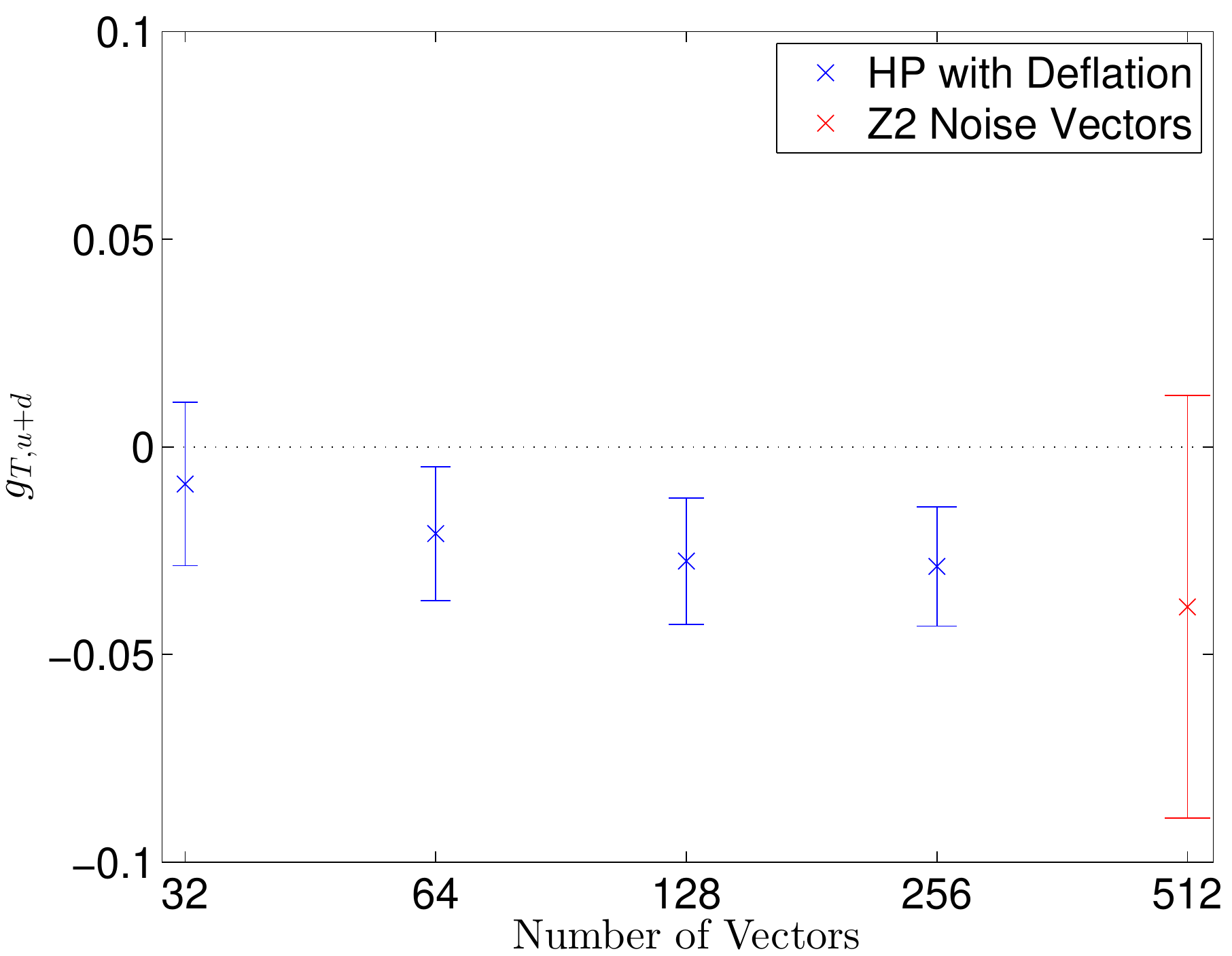}}
\subcaptionbox{$g_V$ 100 cfgs\label{gV_100_comparison}}
{\includegraphics[width=0.45\textwidth]{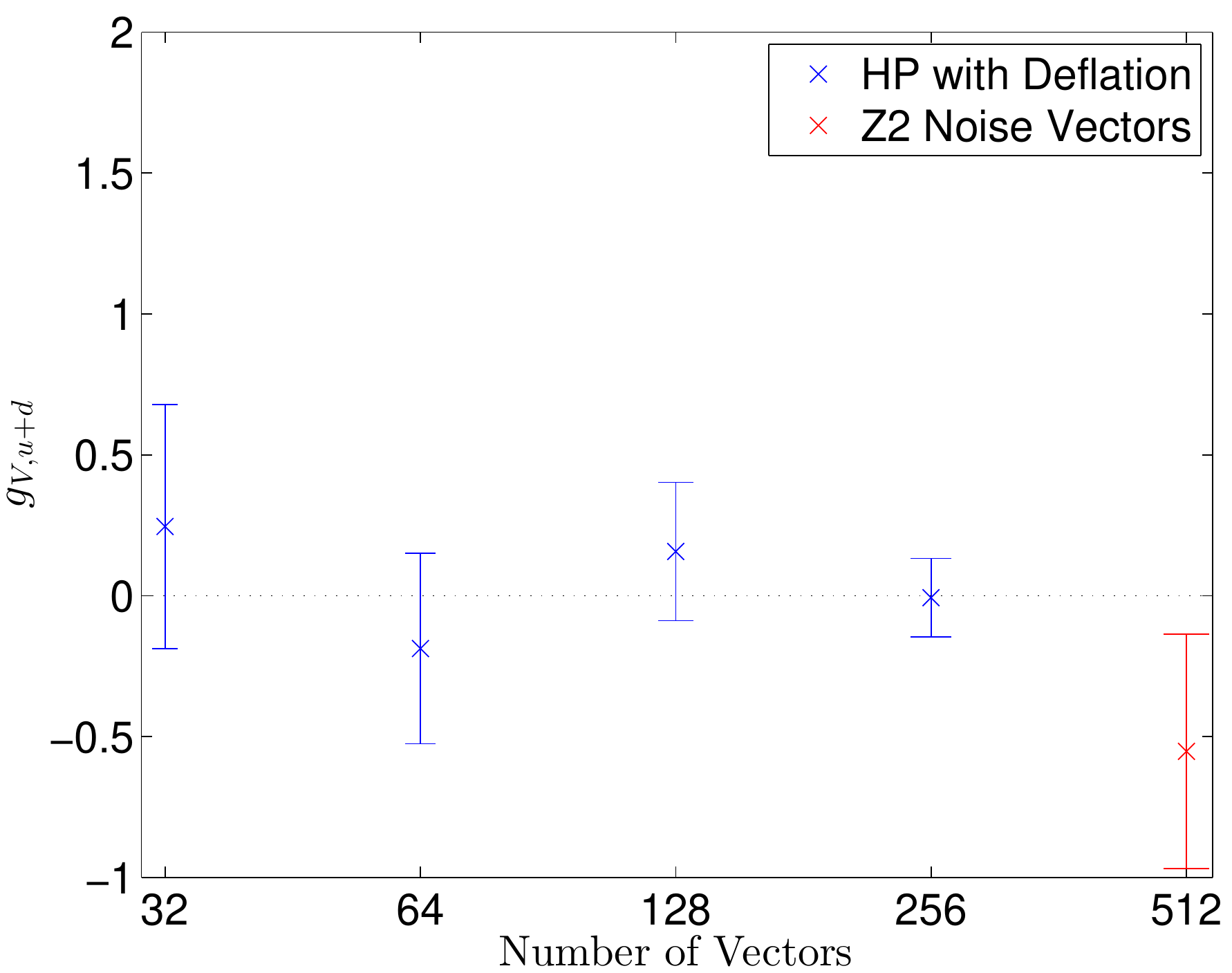}}
\caption{$gA$ $gS$, $gT$, and $gV$ computed with 32, 64, 128, and 256 HP vectors and 500 deflation vectors, compared to the MC estimate 
with 512 $\mathbb{Z}_2$ vectors.}
\end{figure}

The effectiveness of deflated HP in computing disconnected
contributions to nucleon charges is shown in
Figs.~\ref{gA_100_comparison},~\ref{gS_100_comparison},~\ref{gT_100_comparison},
and~\ref{gV_100_comparison}. The source-sink separation in the 3-point
function was $t_{sep}=10$ and the loop insertion is halfway at
$\tau=5$. The two-point correlation functions in this calculation were
computed using the all-mode-averaging method~\cite{Blum:2012uh} with
96 low-precision and 3 high-precision measurements. Deflated HP was
compared to an MC run using $\mathbb{Z}_2$ noise vectors. The MC run
used 512 $Z_2$ noise vectors improved with the
hopping parameter expansion \cite{Foster:1998vw, Michael:1999rs} and
the truncated solver method \cite{Bali:2009hu}.

The data in
Figs.~\ref{gA_100_comparison},~\ref{gS_100_comparison},~\ref{gT_100_comparison},
and~\ref{gV_100_comparison} show that deflated HP with 32
HP vectors performed better than 512 $Z_2$ noise vectors.  The cost of
generating the 500 deflation vectors was approximately $20\%$ of the
cost of the 256 probing vectors. This can be computed from the timings
in Re.~\cite{Gambhir:2016uwp}.  For the case of $g_A$, $g_T$, and
especially $g_V$, substantial error reduction was observed with the
Deflation HP method.  For the scalar charge, $g_S$, fluctuations from
the ensemble average are large and contribute more significantly to
the error. Thus a highly accurate trace per configuration is not
beneficial.

\section{Conclusions and Outlook}

We emphasize the synergy of the two algorithms: Hierarchical Probing
(HP) and Deflation. Although HP's effectiveness decreases at light
quark masses, substantial variance reduction is achieved when
augmented with deflation. Disconnected contributions to the 
three nucleon charges $g_{A,T,V}$ benefit from
our more efficient trace estimator, while little improvement is
observed for $g_S$. We are currently extending our analysis to
isoscalar vector and axial vector form factors: $G_E^s(Q^2)$,
$G_M^s(Q^2)$, $G_A^s(Q^2)$, and $G_S^s(Q^2)$.

\section*{Acknowledgments}
This work has been supported by NSF under grants No. CCF 1218349 and
ACI SI2-SSE 1440700, and by DOE under a grant No. DE-FC02-12ER41890.
KO and AG have been supported by the U.S. Department of Energy through
Grant Number DE- FG02-04ER41302.  KO has been supported through
contract Number DE-AC05-06OR23177 under which JSA operates the Thomas
Jefferson National Accelerator Facility.  AG has been supported by the
U.S. Department of Energy, Office of Science, Office of Workforce
Development for Teachers and Scientists, Office of Science Graduate
Student Research (SCGSR) program. The SCGSR program is administered by
the Oak Ridge Institute for Science and Education for the DOE under
contract number DE-AC05-06OR23100. This research used resources of the
National Energy Research Scientific Computing Center, a DOE Office of
Science User Facility supported by the Office of Science of the
U.S. Department of Energy under Contract
No. DE-AC02-05CH11231. Additionally, this research used resources of
the Oak Ridge Leadership Computing Facility at the Oak Ridge National
Laboratory, which is supported by the Office of Science of the
U.S. Department of Energy under Contract No. DE-AC05-00OR22725. The
work of R.G. and B.Y. is supported by the U.S. Department of Energy,
Office of Science, Office of High Energy Physics under contract number
DE-KA-1401020 and the LANL LDRD program.

\FloatBarrier

\bibliography{qcd}

\begin{thebibliography}{10}

\bibitem{Babich:2010qb}
R.~Babich, J.~Brannick, R.~C. Brower, M.~A. Clark, T.~A. Manteuffel, S.~F.
  McCormick, J.~C. Osborn, and C.~Rebbi.
\newblock {Adaptive multigrid algorithm for the lattice Wilson-Dirac operator}.
\newblock {\em Phys. Rev. Lett.}, 105:201602, 2010.

\bibitem{Babich:2007jg}
R.~Babich, R.~Brower, M.~Clark, G.~Fleming, J.~Osborn, and C.~Rebbi.
\newblock {Strange quark contribution to nucleon form factors}.
\newblock {\em PoS}, LAT2007:139, 2007.

\bibitem{Babich:2011}
R.~Babich, R.~Brower, M.~Clark, G.~Fleming, J.~Osborn, C.~Rebbi, and
  D.~Schaich.
\newblock {Exploring strange nucleon form factors on the lattice}.
\newblock 4 May 2011.

\bibitem{Bali:2009hu}
G.~S. Bali, S.~Collins, and A.~Schaefer.
\newblock {Effective noise reduction techniques for disconnected loops in
  Lattice QCD}.
\newblock (arXiv hep-lat 0910.3970v2), 2010.

\bibitem{Blum:2012uh}
Thomas Blum, Taku Izubuchi, and Eigo Shintani.
\newblock {New class of variance-reduction techniques using lattice
  symmetries}.
\newblock {\em Phys. Rev.}, D88(9):094503, 2013.

\bibitem{Dong:1993pk}
Shao-Jing Dong and Keh-Fei Liu.
\newblock {Stochastic estimation with Z(2) noise}.
\newblock {\em Phys. Lett.}, B328:130--136, 1994.

\bibitem{Foley:2005ac}
J.~Foley., K.~J. Juge, A.~O'Cais, M.~Peardon, S.~Ryan, and J.-I. Skullerud.
\newblock Practical all-to-all propagators for lattice qcd.
\newblock {\em Comput. Phys. Commun.}, 172:145--162, 2005.

\bibitem{Foster:1998vw}
M.~Foster and Christopher Michael.
\newblock {Quark mass dependence of hadron masses from lattice QCD}.
\newblock {\em Phys. Rev.}, D59:074503, 1999.

\bibitem{Gambhir:2016uwp}
Arjun~Singh Gambhir, Andreas Stathopoulos, and Kostas Orginos.
\newblock {Deflation as a Method of Variance Reduction for Estimating the Trace
  of a Matrix Inverse}.
\newblock 2016.

\bibitem{Hutchinson_90}
M.~F. Hutchinson.
\newblock A stochastic estimator of the trace of the influence matrix for
  {L}aplacian smoothing splines.
\newblock {\em J. Commun. Statist. Simula.}, 19(2):433--450, 1990.

\bibitem{Michael:1999rs}
Christopher Michael, M.~S. Foster, and C.~McNeile.
\newblock {Flavor singlet pseudoscalar and scalar mesons}.
\newblock {\em Nucl. Phys. Proc. Suppl.}, 83:185--187, 2000.

\bibitem{Morningstar_Peardon_etal_2011}
C.~Morningstar, J.~Bulava, J.~Foley, K.J. Juge, D.~Lenkner, M.~Peardon, and
  C.H. Wong1.
\newblock Improved stochastic estimation of quark propagation with {L}aplacian
  {H}eaviside smearing in lattice {QCD}.
\newblock {\em Phys. Rev. D}, 83(114505), 2011.

\bibitem{Neff:2001zr}
H.~Neff, N.~Eicker, T.~Lippert, John~W. Negele, and K.~Schilling.
\newblock {On the low fermionic eigenmode dominance in QCD on the lattice}.
\newblock {\em Phys. Rev.}, D64:114509, 2001.

\bibitem{Stathopoulos:2013aci}
Andreas Stathopoulos, Jesse Laeuchli, and Kostas Orginos.
\newblock {Hierarchical probing for estimating the trace of the matrix inverse
  on toroidal lattices}.
\newblock 2013.

\bibitem{Stathopoulos06primme:preconditioned}
Andreas Stathopoulos and James~R. McCombs.
\newblock Primme: Preconditioned iterative multimethod eigensolver: Methods and
  software description, 2006.

\bibitem{Stathopoulos:2007zi}
Andreas Stathopoulos and Kostas Orginos.
\newblock {Computing and deflating eigenvalues while solving multiple right
  hand side linear systems in quantum chromodynamics}.
\newblock {\em SIAM J. Sci. Comput.}, 32:439--462, 2010.

\bibitem{Tang_Saad_traceInv}
J.~Tang and Yousef Saad.
\newblock Domain-decomposition-type methods for computing the diagonal of a
  matrix inverse.
\newblock {\em Report UMSI 2010/114}, (MSI, University of Minnesota).

\end{thebibliography}
\bibliographystyle{plain}

\end{document}